# Decryption Through Polynomial Ambiguity: Noise-Enhanced High-Memory Convolutional Codes for Post-Quantum Cryptography

Meir Ariel

School of Electrical and Computer Engineering, Faculty of Engineering, Tel Aviv University, Tel Aviv, Israel

## ABSTRACT

*We present a novel approach to post-quantum cryptography that employs directed-graph decryption of noise-enhanced high-memory convolutional codes. The proposed construction generates random-like generator matrices that effectively conceal algebraic structure and resist known structural attacks. Security is further reinforced by the deliberate injection of strong noise during decryption, arising from polynomial division: while legitimate recipients retain polynomial-time decoding, adversaries face exponential-time complexity. As a result, the scheme achieves cryptanalytic security margins surpassing those of Classic McEliece by factors exceeding $2^{200}$. Beyond its enhanced security, the method offers greater design flexibility, supporting arbitrary plaintext lengths with linear-time decryption and uniform per-bit computational cost, enabling seamless scalability to long messages. Practical deployment is facilitated by parallel arrays of directed-graph decoders, which identify the correct plaintext through polynomial ambiguity while allowing efficient hardware and software implementations. Altogether, the scheme represents a compelling candidate for robust, scalable, and quantum-resistant public-key cryptography.*

## 1. INTRODUCTION

Code-based cryptography, first proposed by Robert McEliece in 1978 [1], employs binary Goppa codes to construct a public-key encryption scheme. Its central idea is to disguise an error-correcting code through invertible linear transformations, producing a public key that conceals the underlying structure. During encryption, the sender deliberately introduces controlled errors into the codeword, making decryption computationally infeasible for adversaries, while the legitimate recipient—holding the private key—can efficiently correct these errors and recover the original plaintext.

Since its inception, many variants of the McEliece cryptosystem have been proposed, typically by replacing Goppa codes with alternative code families [2–3]. However, most of these alternatives have failed to retain the original scheme's strong security guarantees. As a result, Classic McEliece advanced to Round 4 of the NIST Post-Quantum Cryptography (PQC) standardization process [4], though it was not selected for final adoption. Comprehensive reviews of code-based systems and their role in the PQC landscape are provided in [5–6].

Despite its robustness and long-standing security record, Classic McEliece suffers from practical limitations. Its reliance on fixed Goppa codes imposes relatively rigid error-correction capabilities—restricted to parameter sets such as ($N, K, t$) = (1024, 524, 50), (4096, 3556, 45), and (8192, 6528, 128)—where $K \times N$ denotes the generator matrix dimensions and $t$ the error-correction capacity. These constraints hinder adaptability across different security levels and complicate deployment in diverse application contexts.

To overcome these limitations, our work introduces a novel post-quantum cryptographic framework based on noise-enhanced high-memory convolutional codes. The proposed method extends the McEliece paradigm by enabling dynamic code design, stronger key concealment, and scalable decoding. Its principal advantages include:

**Diverse Code Selection**: A broad family of convolutional codes can serve as public and private keys, enabling customization for desired performance and security levels.

**Enhanced Public Key Security**: The scheme's high-density, random-like generator matrix conceals structural information more effectively than the low-density matrices or those with recognizable patterns, used in other code-based systems [7]. The ability to employ powerful convolutional codes with higher error-correcting capacity permits the deliberate introduction of stronger noise, increasing cryptanalytic resistance by factors exceeding $2^{200}$ (depending on key length) compared with Classic McEliece.

**Scalability and Linear-Time Decoding**: Unlike fixed-dimension block codes, the proposed approach supports plaintexts of arbitrary length. Decoding complexity grows linearly with key length, ensuring scalability without loss of efficiency.

**Efficient Implementation**: The decryption process employs an array of directed-graph decoders to identify the correct plaintext through polynomial ambiguity, while allowing efficient and straightforward hardware and software implementation.

The remainder of this paper is organized as follows: Section 2 describes the encoder design of high-memory convolutional codes. Section 3 details the construction of public and private keys, forming the foundation of the proposed framework. Section 4 outlines the encryption process, and Section 5 presents decryption through polynomial ambiguity. Section 6 analyses potential eavesdropping attacks, while Section 7 discusses polynomial selection strategies and error-rate considerations critical to both security and reliable decryption. Section 8 demonstrates a worked implementation example, and Section 9 evaluates cryptanalytic resistance and computational complexity. Finally, Section 10 concludes with key findings.

## 2. HIGH-MEMORY CONVOLUTIONAL CODES

Throughout this paper, binary vectors are denoted using bold lowercase letters (e.g., $a, b$) with their corresponding polynomial representations written as $a(x), b(x)$. The notations $a$ and $a(x)$ are used interchangeably depending on the context. The Hamming weight of a vector $a$ is denoted by wt($a$). We adopt standard conventions from error-correcting code theory: $m$ denotes an information sequence; $c$ and $d$ represent codewords (for masked and unmasked codes, respectively); $s$ denotes a syndrome; and $r$ represents a CRC polynomial. We use calligraphic letters, such as $\mathcal{D}$ and $\mathcal{L}$ to denote sets of binary vectors. Polynomial generator matrices associated with convolutional codes are denoted by uppercase letters (e.g., $A(x), B(x)$), with their corresponding scalar representations written as $A$ and $B$. Given $n$ vectors:

$$v = (v_0, v_1, v_2, \ldots), u = (u_0, u_1, u_2, \ldots), w = (w_0, w_1, w_2, \ldots), \ldots$$

the *interleaving* operation, denoted $u \curlywedge v \curlywedge w \ldots$, produces a new vector by alternating the elements of the input vectors:

$$(v \curlywedge u \curlywedge w \ldots) = (v_0, u_0, w_0, \ldots v_1, u_1, w_1, \ldots v_2, u_2, w_2, \ldots) \quad (1)$$

To *Deinterleave* an interleaved vector and extract one of its constituent components, we define the following operation:

$$(v \curlywedge u \curlywedge w \ldots)_i \quad (2)$$

This operation extracts the elements located at positions

$$i, n+i, 2n+i, \ldots$$

in the interleaved vector. Specifically,

$$(v \curlywedge u \curlywedge w \ldots)_0 = v$$

Denote by $p_i(x)$ a binary polynomial of memory up to $p$

$$p_i(x) = \sum_{j=0}^{p} a_j x^j, \text{ where } a_j \in \mathbb{F}_2 \quad (3)$$

A Convolutional Code (CC) is defined by a set of $n$ constituent polynomials, customarily structured into the form of a *polynomial generator matrix*, denoted as

$$G_P(x) = [p_0(x), p_1(x), \ldots, p_{n-1}(x)] \quad (4)$$

The matrix $G_P(x)$ determines both the rate and the error-correction capability of the CC, which is typically characterized by its *free distance*, $d_{free}$. The parameter $p$ also determines the number of states, $2^p$, in the trellis diagram—a directed graph representing the code structure—and thus directly affects the decoding complexity when using the Viterbi algorithm. An example of an encoder is shown in Fig. 1. In this example, $n = 2$, and the two polynomials obtained by multiplying the input by $p_0(x)$ and $p_1(x)$, respectively, are interleaved according to Equation (1) to form the encoder output.

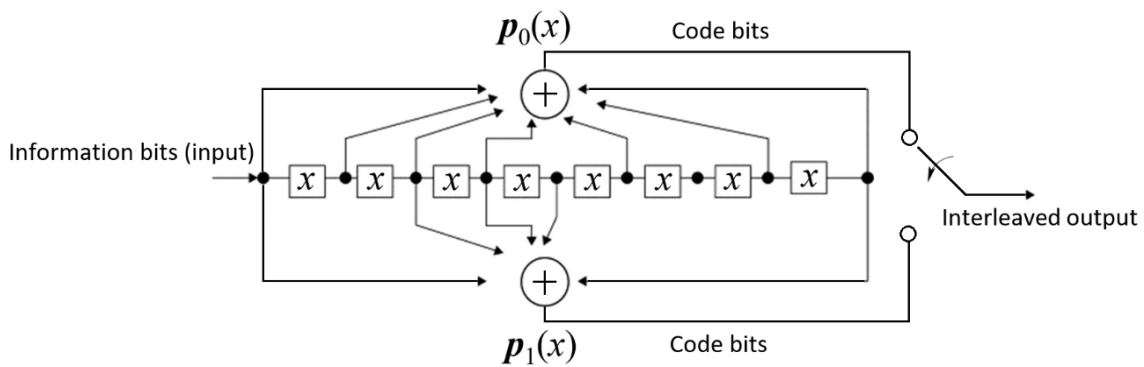

Figure 1. Convolutional encoder for $G_P(x) = [1+x+x^2+x^3+x^5+x^7+x^8, 1+x^2+x^3+x^4+x^8]$

For simplicity, we consider good CCs of rate $1/n$, although the construction presented herein is applicable to any CC, including punctured codes. The matrix $G_P(x)$ has a scalar representation $G_P$ given by:

$$G_P = \begin{bmatrix} g_0 & g_1 & g_2 & \cdots & g_p & 0 & 0 & 0 & \cdots \\ 0 & g_0 & g_1 & \cdots & g_{p-1} & g_p & 0 & 0 & \cdots \\ 0 & 0 & g_0 & \cdots & g_{p-2} & g_{p-1} & g_p & 0 & \cdots \\ \vdots & & & & & & & & \end{bmatrix} \quad (5)$$

where $\boldsymbol{g}_i$ is the $1 \times n$ matrix of coefficients of $x^i$ (in the general case $\boldsymbol{g}_i$ is a $k \times n$ scalar matrix). For example, for $G_P(x) = [1+x^2,\ 1+x+x^2]$ we have $\boldsymbol{g}_0 = [1\ 1]$, $\boldsymbol{g}_1 = [0\ 1]$, $\boldsymbol{g}_2 = [1\ 1]$ and $\boldsymbol{0} = [0\ 0]$.

To each polynomial $\boldsymbol{p}_i(x)$ we associate a *high-memory polynomial* $\boldsymbol{q}_i(x)$ with degree up to $q$,

$$\boldsymbol{q}_i(x) = \sum_{j=0}^{q} b_j x^j,\ \text{where } b_j \in \mathbb{F}_2 \quad (6)$$

The set of these $n$ high-memory polynomials forms the polynomial matrix:

$$G_Q(x) = [\boldsymbol{q}_0(x),\ \boldsymbol{q}_1(x),\ \ldots,\ \boldsymbol{q}_{n-1}(x)] \quad (7)$$

The proposed algorithm imposes no restriction on the choice of $G_P(x)$ and $G_Q(x)$, and their polynomials may be reducible or irreducible. However, in practice the degrees $p$ and $q$ are chosen such that $p + q > 200$ and $p \ll q$. Furthermore, the particular selection of $G_Q(x)$ significantly affects the propagation of the error during decryption (as discussed in Section 7), necessitating an appropriate choice to maintain a tolerable error rate at the decoder. The *high-memory* polynomial generator matrix $G_{PQ}(x)$ is defined as:

$$G_{PQ}(x) = [\boldsymbol{p}_0(x)\boldsymbol{q}_0(x),\ \boldsymbol{p}_1(x)\boldsymbol{q}_1(x),\ \ldots,\ \boldsymbol{p}_{n-1}(x)\boldsymbol{q}_{n-1}(x)] \quad (8)$$

A corresponding finite-dimensional high-memory scalar generator matrix is given by

$$G_{PQ} = \begin{bmatrix} g_0 & g_1 & g_2 & \cdots & g_{p+q} & 0 & \cdots & & & 0 \\ 0 & g_0 & g_1 & \cdots & g_{p+q-1} & g_{p+q} & 0 & \cdots & & 0 \\ 0 & 0 & g_0 & \cdots & g_{p+q-2} & g_{p+q-1} & g_{p+q} & 0 & \cdots & 0 \\ \vdots & & & & & & & & & \\ 0 & 0 & 0 & \cdots & g_0 & g_1 & \cdots & g_{p+q-2} & g_{p+q-1} & g_{p+q} \end{bmatrix} \quad (9)$$

The number $K$ of rows of $G_{PQ}$ and the corresponding number of columns, given by

$$N = n(K + p + q)$$

can be determined by the owner of the key. Multiplying each polynomial of $G_P(x)$ by a corresponding high-memory polynomial in $G_Q(x)$ substantially increases the run-length of each row in $G_{PQ}$, i.e., the length of a sequence within the row that begins and ends with "1."

When employed for error correction, the $K \times N$ matrix $G_{PQ}$ corresponds to a CC with memory length $p + q$, that can be described using a trellis diagram comprising $N/n$ segments and up to $2^{p+q}$ states. A conventional trellis diagram starts from a single state and expands to $2^{p+q}$ states over $p + q$ segments. Additionally, $p + q$ zeroes need be appended at end of the information sequence to drive the trellis to a single state. A Viterbi decoder can be used for maximum-likelihood hard-decision decoding (based on a Hamming distance metric) of the CC. However, this approach becomes impractical for large values of $2^{p+q}$.

Since $G_{PQ}$ has finite dimensions, the corresponding CC exhibits block code structure, defined by a generator matrix with a structured diagonal pattern. Each row is obtained by a right shift of *n* bits relative to the previous row. Although the code description is intricate, permuting the columns of $G_{PQ}$ does not obscure its structure, as the position of zero sequences in each column still reveal discernible column patterns.

## 3. PUBLIC AND PRIVATE KEY CONSTRUCTION

To obfuscate $G_{PQ}$, disrupting its diagonal structure while maximizing the run length, we introduce a *K × N masking matrix $\tilde{G}$ of rank l,* where each row is randomly selected from a predefined set $\mathcal{L}$ of *l* random binary vectors of length *N* and their linear combinations. Denote the linear span of $\mathcal{L}$ as:

$$\text{LS}(\mathcal{L}) = \{\boldsymbol{l}_0, \boldsymbol{l}_1, \ldots, \boldsymbol{l}_{2^l-1}\} \tag{10}$$

where $\boldsymbol{l}_i$, represents a distinct linear combination of the elements of $\mathcal{L}$. The masked generator matrix is then constructed as $G_{PQ} + \tilde{G}$, where addition is performed over $\mathbb{F}_2$. The resulting matrix is now fully dense, having lost its original diagonal structure. Moreover, since $G_{PQ}$ has full rank then $G_{PQ} + \tilde{G}$ can also be chosen to have full rank, making it well suited for subsequent invertible transformations. The final encryption matrix $G$, which serves as the public key, is obtained by applying both row and column transformations to $G_{PQ} + \tilde{G}$:

$$G = S(G_{PQ} + \tilde{G})R \tag{11}$$

where *S* is a random non-singular *K × K* binary matrix used to hide the encoding—that is correspondence between plaintexts (information words) and ciphertexts (codewords), and *R* is a random *N × N* permutation matrix. The resulting random linear code described by $G$ is referred to as a *high-memory masked convolutional code* (MCC).

The transformation from $G_P$ to $G$ involves several steps, some fully reversible and others only semi-reversible, ultimately producing a fully dense, random-like matrix structure. This structure provides significantly stronger resistance to cryptanalysis than low-density matrices, or those exhibiting distinct and easily recognizable patterns.

In the proposed noise-enhanced scheme, encryption deliberately introduces controlled bit-flips to mask the code structure and increase the decoding complexity faced by an attacker. Let *e* denote the probability that a ciphertext bit is flipped. As will be shown later, the parameter *e* also induces additional noise during decryption, further reinforcing security. This parameter is defined by the public-key owner and is chosen to balance reliable decryptability with maximal cryptanalytic strength.

To finalize the construction of the public key, an error-detection mechanism is incorporated using a cyclic redundancy check (CRC) defined by a polynomial *r(x)* of degree *r*. This encoding step, applied to the plaintext prior to the encoding by $G$, enables the recipient to identify potential decoding failures. Such detection is essential because, unlike block codes, CCs do not guarantee successful decoding, even when the actual error rate remains within theoretically correctable limits.

We are now ready to define the *public key* as:

$$\{G, e, \boldsymbol{r}\} \tag{12}$$

and the *private key* as:

$$\{S, R, G_P(x), G_Q(x), \tilde{G}\}. \tag{13}$$

At the decoder, the inverse permutation must first be applied, and the effect of the masking matrix $\tilde{G}$ must then be reversed. However, this masking operation is not entirely reversible. Although the recipient possesses complete knowledge of $\tilde{G}$, the plaintext itself remains unknown. Consequently, the specific linear combination of the rows of $\tilde{G}$ that was added during encryption cannot be uniquely reconstructed. The procedure used to resolve this *polynomial ambiguity* is described in Section 5.

Furthermore, the obfuscation induced by the high-memory polynomials $G_Q(x)$ must be addressed before decoding. A trellis decoder of feasible complexity can operate only on the generator matrix $G_P$, not on its high-memory variant $G_{PQ}$. Attempting to invert multiplication by the high-memory polynomials in the presence of noise is inherently difficult, as it can trigger error propagation and increase the likelihood of decryption failures at the recipient. At the same time, this very error propagation amplifies the adversary's decoding decryption complexity, thereby strengthening security. A judicious choice of $G_Q(x)$ can balance these opposing effects—limiting error propagation while maintaining the desired security margin.

## 4. ENCRYPTION BY SENDER

Assume that the sender possesses the public key. The following steps are performed to generate the ciphertext:

**Step1** - **Generation of Plaintext:**

A random plaintext *m* of length *K* - *r* bits is generated, and its polynomial representation is denoted by *m*(*x*).

**Step 2** - **Appending CRC:**

To append *r* CRC bits to *m*, the following procedure is used:

1. Multiply *m*(*x*) by $x^r$ (effectively shifting it).
2. Divide $x^r$*m*(*x*) by the polynomial *r*(*x*) and compute the remainder.
3. Append the remainder to *m* to form a binary vector of length *K*, denoted as $m_r$.

**Step 3** - **Codeword Generation**

The codeword *c* of length *N* is computed as:

$$c = m_r G. \tag{14}$$

**Step 4** - **Error Introduction:**

Random errors are introduced to *c* by flipping each bit with probability *e*. The actual number of errors injected by the sender is random and unknown to the recipient. The resulting ciphertext is given by

International Journal on Cryptography and Information Security (IJCIS), Vol. 15, No.1/2/3/4, December 2025

$$c_e = c + e \tag{15}$$

where $e$ is the random error vector generated by the sender.

**Step 5** - **Transmission:**

The ciphertext $c_e$ is then transmitted by the sender to the recipient.

It should be noted that, since $e$ is generated randomly, it can contain more than *eN* errors or form localized clusters thereby increasing the risk of decoding failure. Furthermore, polynomial division at the decoder amplifies the effective error weight, contributing to additional obfuscation. Nevertheless, owing to the strong error-correcting capability of the CC defined by $G_P$, the probability of decoding failure remains very low even for relatively large values of *e*, as demonstrated in Section 9. In practical implementations, rare decoding failures are reliably detected by either the CRC or, alternatively—as described in Section 7—by comparing the actual number of corrected errors with the estimated value. This enables the recipient to select the next most likely plaintext candidate or request a retransmission when necessary.

## 5. DECRYPTION BY THE RECIPIENT

The recipient possesses the public and private keys along with the received ciphertext, i.e.,

$$\{S, R, G_P(x), G_Q(x), \tilde{G}, G, e, r, c_e\} \tag{16}$$

The following steps are performed by the recipient to decrypt the ciphertext:

**Step 1** - **Inverse Permutation:**

Apply the inverse permutation to $c_e$ to obtain $\tilde{c}$. (Note that if $R$ is a permutation matrix, then $R^{-1} = R^T$). Thus,

$$\tilde{c} = c_e R^T \tag{17}$$

Since $e$ is a random error vector with an unknown Hamming weight,

$$eR^T = c_e R^T - cR^T \tag{18}$$

is simply another random error vector with the same weight, i.e.,

$$\text{wt}(eR^T) = \text{wt}(e) \tag{19}$$

**Step 2** - **Unmasking:**

Reverse the masking introduced by the summation of $\tilde{G}$ and $G_{PQ}$. The vector $\tilde{c}$ can be expressed as:

$$\tilde{c} = m_r S G_{PQ} + m_r S \tilde{G} = m_r S G_{PQ} + l_i, \text{ with } l_i \in \text{LS}(\mathcal{L}) \tag{20}$$

However, although $\tilde{G}$ is known to the recipient, the value of $l_i$ is unknown and may be any of the $2^l$ members of LS($\mathcal{L}$). Therefore, the set of possible unmasked vectors, denoted by $\mathcal{M}$, is given by:

$$\mathcal{M} = \{\tilde{c} - l_i \mid l_i \in \mathrm{LS}(\mathcal{L})\} \tag{21}$$

**Step 3** - **Inverting the High-Memory Polynomial Multiplication:**

The next step involves reversing the multiplication by the high-memory polynomials $G_Q(x)$. Using polynomial representations, each member of $\mathcal{M}$ can be regarded as the result of interleaving the following $n$ polynomials, each of length $N/n$ :

$$(\tilde{c} - l_i)_0 \wedge (\tilde{c} - l_i)_1 \wedge \ldots \wedge (\tilde{c} - l_i)_{n-1} \tag{22}$$

where the elements of $(\tilde{c} - l_i)_j$ appear at the

$$j\text{th}, (n + j)\text{th}, (2n + j)\text{th}, \ldots, (\tfrac{N}{n} - 1 + j)\text{th}$$

positions of the interleaved vector $\tilde{c} - l_i$. Therefore, each such polynomial $(\tilde{c} - l_i)_j$, shall be divided by the corresponding high-memory polynomial $q_j$ to obtain a quotient polynomial, denoted by $(d_i)_j$

$$(d_i)_j = \frac{(\tilde{c} - l_i)_j}{q_j} \tag{23}$$

The value of $(d_i)_j$ naturally depends on the value of $l_i$. At this stage of decryption, the recipient cannot determine which quotient, $(d_i)_j$, is the correct one for index $j$. Furthermore, since $(\tilde{c} - l_i)_j$ may contain errors, the division might yield non-zero remainders. These remainders can either be ignored or be detected and subtracted if $q_j$ is carefully chosen to function as an error-detecting code. When ignored, we assume that errors left in the quotient will be corrected by the Viterbi decoder in Step 5 below.

**Step 4** - **Interleaving of Quotients:**

The quotients are re-interleaved to form a vector $d_i$:

$$d_i = (d_i)_0 \wedge (d_i)_1 \wedge \ldots \tag{24}$$

In the general case, there are no more than $2^l$ variants in the set of interleaved quotients, denoted as

$$\mathfrak{D} = \{d_i\}_{i=0}^{2^l - 1} \tag{25}$$

All candidates in $\mathfrak{D}$ need be considered for decryption.

**Step 5** - **Parallel Viterbi Decoding:**

We now aim to determine $\hat{m}_r$, the most likely value of $m_r$ (i.e., the plaintext with the appended CRC), by applying Viterbi decoding to all possible members of the set $\mathfrak{D}$. Since $\mathfrak{D}$ contains $2^l$ candidate vectors, these directed-graph decoding can be performed in parallel for improved efficiency. The most likely codeword $\hat{d}$ is determined by

$$\hat{d} = d_i - \hat{e} \tag{26}$$

where $\hat{e}$ is the error vector with the minimum Hamming weight among the outcomes of all $2^l$ parallel Viterbi decoders. The index $i$ in Equation (26) corresponds to the decoder that processes the correct $d_i$. The same Viterbi decoder also reveals $\hat{m}_r S$, the transformed plaintext with appended CRC that generated $\hat{d}$.

Importantly, a key distinction exists among the $2^l$ members of $\mathfrak{D}$. For exactly one variant, say $d_i$, Viterbi decoding will yield a codeword at a Hamming distance of approximately $eN + \alpha$ from $d_i$, where $eN$ is the injected error weight, and α denotes the number of additional errors introduced by the polynomial division process. The value of α can be estimated via simulations, as described in Section 7. By contrast, decoding any of the other $2^l$ -1 candidates in $\mathfrak{D}$ typically produces codewords at substantially larger Hamming distance than $eN + \alpha$. This larger Hamming distance can be estimated from the CC's weight spectrum, using the Gilbert bound analysis presented in Section 7.

This distinction arises because subtracting an incorrect masking vector from $\tilde{c}$ effectively introduces random errors into the candidate $d_i$, allowing the decoder to reliably identify the correct candidate among all alternatives.

**Step 6: Plaintext Recovery**

To recover the original plaintext, we reverse the transformation induced by the matrix $S$:

$$\hat{m}_r = \hat{m}_r S S^{-1} \qquad (27)$$

Next, the remainder obtained from dividing $\hat{m}_r(x)$ by $r(x)$ must be computed. If this remainder is zero, we declare that

$$m_r = \hat{m}_r \qquad (28)$$

and recover the plaintext $m$ by discarding the $r$ CRC bits from $m_r$. If the remainder is nonzero, the selected codeword $\hat{d}$ is rejected, and the process continues iteratively with the next most likely candidate. This process is repeated until a valid plaintext is identified or until all candidates are exhausted. If no valid plaintext is found, a retransmission of another ciphertext is requested.

Finally, we remark that the structure of the matrix $G_P$ corresponds to a trellis that terminates in a single state. As a result, the Viterbi decoder returns the most likely information sequence padded with $p$ trailing zeros. To recover the original plaintext, theses appended zeroes must be removed from the decoded output. This step ensures that the final recovered message accurately reflects the original input.

The encryption and decryption algorithms are illustrated in the block diagram of Fig. 2.

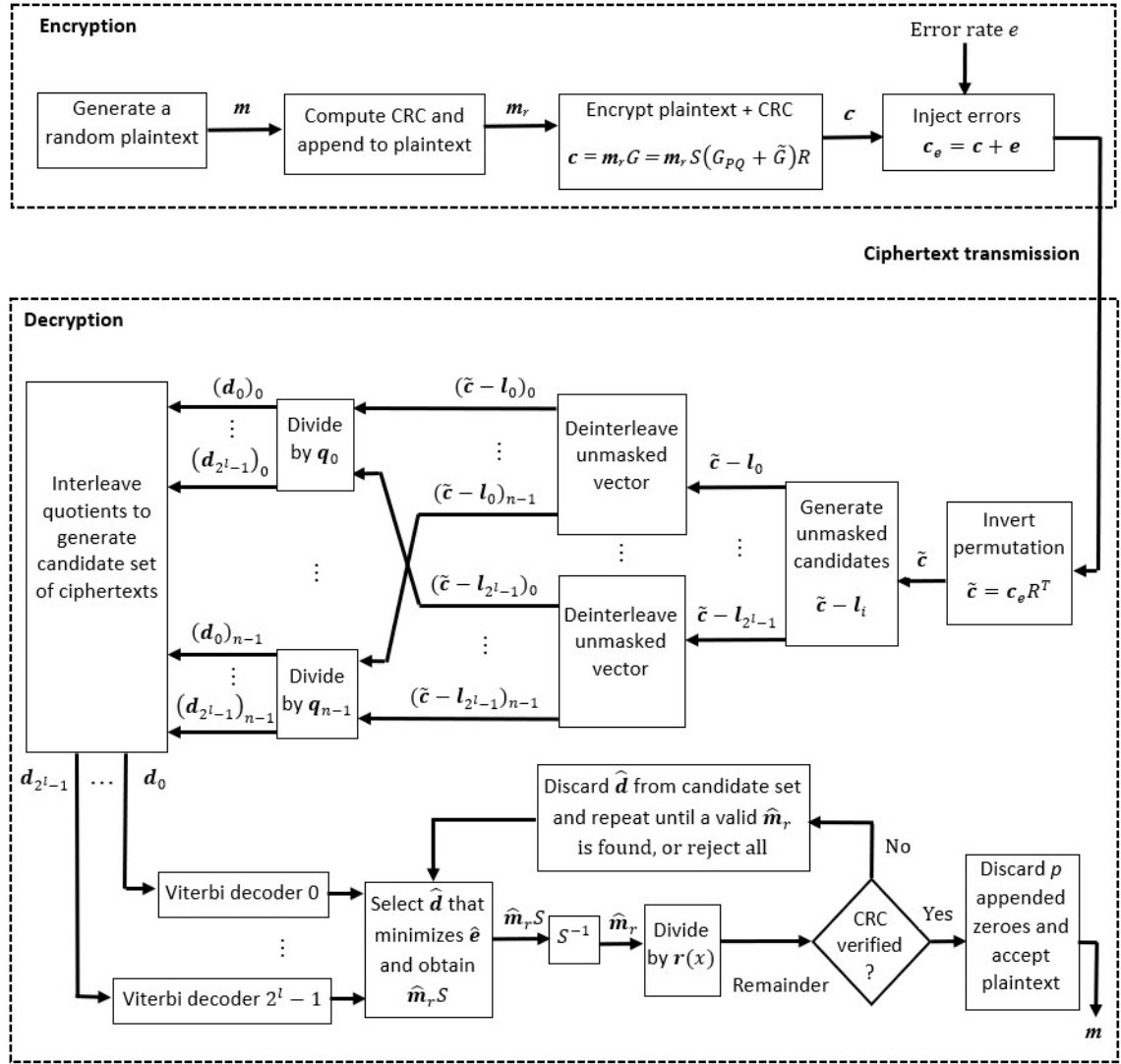

Figure 2. Encryption and decryption block diagram

## 6. SECURITY AGAINST STRUCTURAL ATTACKS

An eavesdropper who observes the public key $G$ and the intercepted ciphertext $c_e$ cannot recover the plaintext $m$ by directly inverting $G$, because $c_e$ contains injected errors. The attacker is therefore constrained to two options: (i) enumerate all possible error patterns—an operation whose complexity grows exponentially with the error weight (which may be of the order of hundreds for ciphertexts several thousand bit long), or (ii) attempt to recover the structured matrix $G_{PQ}$ from the public key $G$. The second approach is defeated by heavy obfuscation implemented by three complementary transformations:

- **Dense masking.** A carefully constructed masking matrix $\tilde{G}$ whose rows are random with Hamming weight close to *N/2*; each row and column is statistically balanced so algebraic structure is hidden.

- **Random row operations.** A nonsingular matrix $S$ scrambles the code basis via invertible linear combinations of rows.

- **Random column permutation.** A permutation matrix $R$ uniformly permutes coordinate positions, destroying any positional patterns.

**Statistical Randomness of the Public Generator Matrix**

Proving that a masked generator matrix is statistically indistinguishable from that of a random linear code remains notoriously difficult. Even for Classic McEliece, no formal proof establishes that (permuted) Goppa codes behave like random linear codes; instead, this assumption has withstood decades of unsuccessful structural attacks at the chosen parameters, and—practically—ISD remains the dominant attack class for those parameter sets. Against this backdrop, we present strong empirical and analytical evidence that the public matrix $G = S(G_{PQ} + \tilde{G})R$ exhibits all statistical properties expected of a random full-rank $(N, K)$ code.

**Rank Preservation:** Consider $G_{PQ} \in \mathbb{F}_2^{K \times N}$ of full row rank $K$, and a dense masking matrix $\tilde{G}$ of rank $l$. The sum $G_{PQ} + \tilde{G}$ therefore has rank at least $K - l$. Because the $l$ fixed random row patterns of $\tilde{G}$ are dense and mutually distinct, the sum $G_{PQ} + \tilde{G}$ will in general have full rank $K$. Moreover, knowledge of the diagonal structure of $G_{PQ}$ enables selection of masking rows in $\tilde{G}$ to guarantee this property. Multiplication by the nonsingular matrix $S$ and permutation by $R$ preserve rank, ensuring that the public key $G$ remains a full-rank generator matrix.

**Column Randomness via Linear Tests:** Let $y_j$ denote the $j$th column of $S(G_{PQ} + \tilde{G})$. Since the rows of $G_{PQ} + \tilde{G}$ are dense with Hamming weight approximately $N/2$, and are further scrambled by the random nonsingular matrix $S$, the entries of $y_j$ behave like independent Bernoulli(1/2) variables. Consequently, the column weights of $S(G_{PQ} + \tilde{G})$ concentrate sharply around $K/2$, matching the expected behaviour of a uniformly random full-rank code.

A *linear test* defined by a nonzero vector $h \in \mathbb{F}_2^K$ computes the parity $h^T y_j \in \{0,1\}$, which equals zero when the overlap between $h$ and $y_j$ has even Hamming weight, and one otherwise. Any persistent bias would require statistical alignment between $h$ and specific row patterns of $\tilde{G}$. However, multiplication by $S$ disperses each fixed row pattern across all coordinates, destroying such alignments. Hence, the outcomes of linear tests are asymptotically uniform, and the empirical column statistics are indistinguishable from those of a full-rank random $(N, K)$ block code, up to deviations of order $2^{-l}$. For practical values of $l$, this deviation is negligible compared with natural sampling fluctuations.

**Suppression of Low-Weight Dual Codewords:** A vector $w \in \mathbb{F}_2^K \setminus \{0\}$ is a *dual vector* if

$$w^T(G_{PQ} + \tilde{G}) = 0.$$

Since $G_{PQ}$ has full rank, $w^T G_{PQ} \neq 0$ for all $w \neq 0$. The equation $w^T G_{PQ} = w^T \tilde{G}$ therefore requires that the linear combination of the fixed dense row patterns specified by $w$ exactly reproduce the nonzero target vector $w^T G_{PQ}$. For dense and mutually distinct patterns, the probability of such alignment for any low-weight $w$ is exponentially small. After multiplication by $S$ and permutation $R$, any rare alignment becomes uncorrelated with coordinate positions. Thus, the number and distribution of low-weight duals in $G$ match those expected in a random linear code.

**Masking Entropy:** A critical asymmetry arises from the masking matrix $\tilde{G}$: while the legitimate recipient known its structure, the adversary does not. Even though its rank $l$ is small, the space of possible masks is astronomically large. Each of the $K$ rows can be chosen independently from $l$ dense random patterns (or their linear combinations), yielding at least $l^K$ distinct possible masks.

For example, with $K = 2000$ and $l = 5$, the number of distinct candidates is $5^{2000}$, corresponding to about 4644 bits of entropy. Exhaustive search over this space is infeasible in practice, and ciphertext errors completely prevent efficient verification of guessed masks.

In contrast, the legitimate recipient who constructed $\tilde{G}$ only needs to consider $2^l$ mask candidates, and can perform polynomial-time parallel decoding, selecting the codeword of minimal Hamming distance. This exponential-versus-polynomial gap is a fundamental source of the security argument.

In conclusion, the obfuscations $S(G_{PQ} + \tilde{G})R$ effectively erase all structural fingerprints of $G_{PQ}$. The resulting public key $G$ exhibits full rank, statistically balanced columns, an absence of low-weight duals, and uniform parity distributions—making it computationally and statistically indistinguishable from a random $(N, K)$ block code. Given the injected ciphertext errors, the only known generic decryption method remains Information-Set Decoding (ISD), whose complexity grows exponentially with the error weight.

**Attack from a Known Convolutional Code:** Even if the attacker knew the exact underlying high-memory CC that generated $G_{PQ}$, direct decoding would remain infeasible. In the public construction, the code is convolved with high-degree polynomials—effectively smearing the run-length and dispersing structure—and then corrupted by the dense matrix and the injected ciphertext errors. The legitimate recipient can invert the polynomial multiplication, reducing the problem to an array of compact trellises on which decoding is tractable; this inversion, however, also reinterprets error patterns in a way that only the intended decoding procedure (with knowledge of $\tilde{G}, S, R$ and the high-degree polynomial factors) can resolve. An attacker who lacks this secret information must simultaneously (i) undo the permutation $R$, (ii) recover or remove the dense mask $\tilde{G}$, and (iii) reverse the high-degree polynomial multiplication—all while contending with amplified and propagated errors. These coupled tasks prevent construction of a useful compact trellis for trellis-based attacks. Consequently, the adversary is left with an effectively random linear code for which the only generic decoding approach is ISD, whose complexity is exponential in the error weight.

## 7. POLYNOMIAL SELECTION AND EXPECTED ERROR RATE

The decryption algorithm involves polynomial divisions that can cause error propagation; a single error in the polynomials $(\tilde{c} - l_i)_j$ may lead to multiple errors in the quotient. To reduce this risk, the polynomial matrix $G_P(x)$ is selected to ensure that the corresponding CC has strong error-correction capabilities (e.g., with free distance $d_{\text{free}} > 20$), outperforming Goppa codes or similar linear block codes with comparable size. This results in rare decoding failures, even at relatively high error rates $e$ (a larger $e$ also implies improved security).

To further limit the probability of error propagation, the polynomials $G_Q(x)$ should be chosen to limit the spread of isolated errors in $(\tilde{c} - l_i)_j$, confining them to just a few errors in the quotient. Given an error rate $e$, simulation can test $G_Q(x)$ by measuring the number of quotient errors for various random error vectors with the same $e$, allowing for an informed selection of the CC.

Let $e_j$ be a random error vector of length $N/n$ with error rate $e$. The total number of additional errors introduced by polynomial division, denoted α, is given by:

$$\alpha = \sum_{j=0}^{n-1}[\text{wt}\left(\frac{e_j}{q_j}\right) - \text{wt}(e_j)] \tag{29}$$

Here, the term $\frac{e_j}{q_j}$ corresponds to the quotient obtained during polynomial division of an error vector of length *N/n* by the divisor polynomial. The expression in Equation (29) captures the cumulative increase in Hamming weight due to the spread of errors in the quotient domain.

A practical way to limit the spread of isolated errors during polynomial division is to choose $\boldsymbol{q}_j$ as a sparse polynomial with widely spaced nonzero exponents. For instance, the trivial choice $\boldsymbol{q}_j(x) = x^i$ (with $(i \leq q)$, completely avoids error propagation, although it contributes minimally to security. A more effective option is to use a two-term polynomial, which both extends the run-length of each row while minimizing error spread:

$$\boldsymbol{q}_j(x) = 1 + x^q \tag{30}$$

Increasing the number of nonzero elements in $\boldsymbol{q}_j(x)$ can dramatically amplify the difference

$$\text{wt}\left(\frac{e_j}{q_j}\right) - \text{wt}(\boldsymbol{e}_j) \tag{31}$$

Therefore, the polynomials of $G_Q(x)$ should be selected such that the resulting estimated error rate

$$\frac{eN+\alpha}{N} \tag{32}$$

remains within the CC's decoding capacity.

Consider a candidate vector $\boldsymbol{d}_i$ defined according to Equation (24), and assume that $\boldsymbol{d}_i$ is an incorrect candidate (i.e., it was incorrectly demasked). We now estimate the Hamming distance between $\boldsymbol{d}_i$ and the closest codeword in the CC. Maximum-likelihood hard-decision Viterbi decoding of a random vector such as $\boldsymbol{d}_i$ effectively returns the coset leader—the error vector that produces the nearest codeword in Hamming distance.

The expected weight of a random coset leader can be approximated using the Gilbert bound, which provides a good estimate of the number of errors corrected in a random vector. For essentially any "good" binary linear code of rate $\rho = K/N$, the typical nearest-neighbour distance $\delta$ is well approximated by the solution of the volume (Gilbert) equation

$$\sum_{i=0}^{\delta} \binom{N}{i} \approx 2^{N(1-\rho)} \tag{33}$$

Using the entropy approximation

$$\sum_{i \leq \delta} \binom{N}{i} \approx 2^{NE(\delta/N)} \tag{34}$$

we obtain

$$E(\delta/N) \approx 1 - \rho \tag{35}$$

where is $E(\cdot)$ is the binary entropy function (base-2). For example, at $\rho = \frac{1}{2}$ we have

$$E\left(\frac{\delta}{N}\right) \approx \frac{1}{2} \implies \frac{\delta}{N} \approx 0.11$$

Similarly, at code rates $\rho = \frac{1}{3}$ and $\rho = \frac{1}{4}$ we obtain $\frac{\delta}{N} \approx 0.174$ and $\frac{\delta}{N} \approx 0.215$, respectively.

## 8. WORKED EXAMPLE

Define the following CC with

$$G_P(x) = [\boldsymbol{p}_0(x), \boldsymbol{p}_1(x)] = [1+x^2, 1+x+x^2].$$

Suppose that a randomly selected plaintext of length 6 is given by $\boldsymbol{m} = [1\ 1\ 1\ 0\ 0\ 1]$, in polynomial form

$$\boldsymbol{m}(x) = 1+x+x^2+x^5.$$

In this example we skip the trivial step of CRC construction, assuming it is already contained within $\boldsymbol{m}(x)$.

A codeword $\boldsymbol{d}(x)$ of the CC is obtained by either interleaving $\boldsymbol{m}(x)\boldsymbol{p}_0(x)$ with $\boldsymbol{m}(x)\boldsymbol{p}_1(x)$ or simply by employing a scalar generator matrix with 6 rows (corresponding to the length of $\boldsymbol{m}$) constructed according to Equation (5)

$$G_P = \begin{bmatrix} 1&1&0&1&1&1&0&0&0&0&0&0&0&0 \\ 0&0&1&1&0&1&1&1&0&0&0&0&0&0 \\ 0&0&0&0&1&1&0&1&1&1&0&0&0&0 \\ 0&0&0&0&0&0&1&1&0&1&1&1&0&0 \\ 0&0&0&0&0&0&0&0&1&1&0&1&1&1&0&0 \\ 0&0&0&0&0&0&0&0&0&0&1&1&0&1&1&1 \end{bmatrix}$$

The codeword $\boldsymbol{d}$ in vector form is given by

$$\boldsymbol{d} = \boldsymbol{m}G_P = [1\ 1\ 1\ 0\ 0\ 1\ 1\ 0\ 1\ 1\ 1\ 0\ 1\ 1]$$

(which matches the interleaved polynomial-wise result). Next, choosing the polynomial matrix

$$G_Q(x) = [1+x^7, x^7]$$

we obtain the corresponding high-memory generator matrix

$$G_{PQ}(x) = [\boldsymbol{p}_0(x)\boldsymbol{q}_0(x), \boldsymbol{p}_1(x)\boldsymbol{q}_1(x)] = [(1+x^2)(1+x^7), (1+x+x^2)(x^7)]$$
$$= [1+x^2+x^7+x^9, x^7+x^8+x^9].$$

Using $G_{PQ}(x)$, we construct the high-memory scalar generator matrix $G_{PQ}$ as in Equation (9) with:

$$\boldsymbol{g}_0 = [10],\ \boldsymbol{g}_1 = [00],\ \boldsymbol{g}_2 = [10],\ \boldsymbol{g}_3 = [00],\ \boldsymbol{g}_4 = [00],\ \boldsymbol{g}_5 = [00],$$
$$\boldsymbol{g}_6 = [00],\ \boldsymbol{g}_7 = [11],\ \boldsymbol{g}_8 = [01],\ \boldsymbol{g}_9 = [11],\ \boldsymbol{0} = [00].$$

The resulting matrix is:

International Journal on Cryptography and Information Security (IJCIS), Vol. 15, No.1/2/3/4, December 2025

$$G_{PQ} = \begin{bmatrix} 1 & 0 & 0 & 0 & 1 & 0 & 0 & 0 & 0 & 0 & 0 & 0 & 0 & 1 & 1 & 0 & 1 & 1 & 1 & 0 & 0 & 0 & 0 & 0 & 0 & 0 & 0 & 0 & 0 & 0 \\ 0 & 0 & 1 & 0 & 0 & 0 & 1 & 0 & 0 & 0 & 0 & 0 & 0 & 0 & 0 & 1 & 1 & 0 & 1 & 1 & 1 & 0 & 0 & 0 & 0 & 0 & 0 & 0 & 0 & 0 \\ 0 & 0 & 0 & 0 & 1 & 0 & 0 & 0 & 1 & 0 & 0 & 0 & 0 & 0 & 0 & 0 & 0 & 1 & 1 & 0 & 1 & 1 & 1 & 0 & 0 & 0 & 0 & 0 & 0 & 0 \\ 0 & 0 & 0 & 0 & 0 & 0 & 1 & 0 & 0 & 0 & 1 & 0 & 0 & 0 & 0 & 0 & 0 & 0 & 0 & 1 & 1 & 0 & 1 & 1 & 1 & 0 & 0 & 0 & 0 & 0 \\ 0 & 0 & 0 & 0 & 0 & 0 & 0 & 0 & 1 & 0 & 0 & 0 & 1 & 0 & 0 & 0 & 0 & 0 & 0 & 0 & 0 & 1 & 1 & 0 & 1 & 1 & 1 & 0 & 0 & 0 \\ 0 & 0 & 0 & 0 & 0 & 0 & 0 & 0 & 0 & 0 & 1 & 0 & 0 & 0 & 1 & 0 & 0 & 0 & 0 & 0 & 0 & 0 & 0 & 1 & 1 & 0 & 1 & 1 & 1 \end{bmatrix}.$$

Note that the run-length of each row has increased from 6 to 20. We then construct $\tilde{G}$ by drawing its rows at random from LS($\mathcal{L}$), where $\mathcal{L}$ is a set of $l$ random vectors of length 30. For simplicity of exposition take

$\mathcal{L} = \{(101010101010101010101010101010), (010101010101010101010101010101)\}$

i.e. the two length-30 alternating patters. Then

$$\text{LS}(\mathcal{L}) = \{(1010\ldots 10), (0101\ldots 01), \mathbf{0}, \mathbf{1}\}$$

where $\mathbf{0}$ and $\mathbf{1}$ are the all-zero and all-one vectors, respectively. The alternating vectors have Hamming weight 15 (i.e., $N/2$), ensuring statistical balance. Choose a dense matrix $\tilde{G}$

$$\tilde{G} = \begin{bmatrix} 1 & 0 & 1 & 0 & 1 & 0 & 1 & 0 & 1 & 0 & 1 & 0 & 1 & 0 & 1 & 0 & 1 & 0 & 1 & 0 & 1 & 0 & 1 & 0 & 1 & 0 & 1 & 0 & 1 & 0 \\ 0 & 1 & 0 & 1 & 0 & 1 & 0 & 1 & 0 & 1 & 0 & 1 & 0 & 1 & 0 & 1 & 0 & 1 & 0 & 1 & 0 & 1 & 0 & 1 & 0 & 1 & 0 & 1 & 0 & 1 \\ 1 & 0 & 1 & 0 & 1 & 0 & 1 & 0 & 1 & 0 & 1 & 0 & 1 & 0 & 1 & 0 & 1 & 0 & 1 & 0 & 1 & 0 & 1 & 0 & 1 & 0 & 1 & 0 & 1 & 0 \\ 0 & 1 & 0 & 1 & 0 & 1 & 0 & 1 & 0 & 1 & 0 & 1 & 0 & 1 & 0 & 1 & 0 & 1 & 0 & 1 & 0 & 1 & 0 & 1 & 0 & 1 & 0 & 1 & 0 & 1 \\ 0 & 1 & 0 & 1 & 0 & 1 & 0 & 1 & 0 & 1 & 0 & 1 & 0 & 1 & 0 & 1 & 0 & 1 & 0 & 1 & 0 & 1 & 0 & 1 & 0 & 1 & 0 & 1 & 0 & 1 \\ 1 & 0 & 1 & 0 & 1 & 0 & 1 & 0 & 1 & 0 & 1 & 0 & 1 & 0 & 1 & 0 & 1 & 0 & 1 & 0 & 1 & 0 & 1 & 0 & 1 & 0 & 1 & 0 & 1 & 0 \end{bmatrix}.$$

then sum $\tilde{G}$ and $G_{PQ}$ to obtain

$$G_{PQ} + \tilde{G} = \begin{bmatrix} 0 & 0 & 1 & 0 & 0 & 0 & 1 & 0 & 1 & 0 & 1 & 0 & 1 & 0 & 0 & 1 & 1 & 1 & 0 & 1 & 1 & 0 & 1 & 0 & 1 & 0 & 1 & 0 & 1 & 0 \\ 0 & 1 & 1 & 1 & 0 & 1 & 1 & 1 & 0 & 1 & 0 & 1 & 0 & 1 & 0 & 1 & 1 & 0 & 0 & 0 & 1 & 0 & 0 & 1 & 0 & 1 & 0 & 1 & 0 & 1 \\ 1 & 0 & 1 & 0 & 0 & 0 & 1 & 0 & 0 & 0 & 1 & 0 & 1 & 0 & 1 & 0 & 1 & 0 & 0 & 1 & 1 & 1 & 0 & 1 & 1 & 0 & 1 & 0 & 1 & 0 \\ 0 & 1 & 0 & 1 & 0 & 1 & 1 & 1 & 0 & 1 & 1 & 1 & 0 & 1 & 0 & 1 & 0 & 1 & 0 & 1 & 1 & 0 & 0 & 0 & 1 & 0 & 0 & 1 & 0 & 1 \\ 0 & 1 & 0 & 1 & 0 & 1 & 0 & 1 & 1 & 1 & 0 & 1 & 1 & 1 & 0 & 1 & 0 & 1 & 0 & 1 & 0 & 1 & 1 & 0 & 0 & 0 & 1 & 0 & 0 & 1 \\ 1 & 0 & 1 & 0 & 1 & 0 & 1 & 0 & 1 & 0 & 0 & 0 & 1 & 0 & 0 & 0 & 1 & 0 & 1 & 0 & 1 & 0 & 1 & 0 & 0 & 1 & 1 & 1 & 0 & 1 \end{bmatrix}$$

This sum is high-density: nearly half of the entries are 1, and the row and column weights concentrate around $K/2$ and $N/2$, respectively. Let

$$S = \begin{bmatrix} 1 & 0 & 0 & 1 & 0 & 0 \\ 0 & 1 & 0 & 0 & 0 & 1 \\ 0 & 0 & 1 & 0 & 0 & 0 \\ 0 & 0 & 1 & 1 & 1 & 0 \\ 0 & 0 & 0 & 0 & 1 & 0 \\ 0 & 0 & 1 & 0 & 1 & 1 \end{bmatrix}$$

Left multiplication by the nonsingular matrix $S$ yields

$$S(G_{PQ} + \tilde{G}) = \begin{bmatrix} 0\ 1\ 1\ 1\ 0\ 1\ 0\ 1\ 1\ 1\ 0\ 1\ 1\ 1\ 0\ 0\ 1\ 0\ 0\ 0\ 0\ 0\ 1\ 0\ 0\ 0\ 1\ 1\ 1\ 1 \\ 1\ 1\ 0\ 1\ 1\ 1\ 0\ 1\ 1\ 1\ 0\ 1\ 1\ 1\ 0\ 1\ 0\ 0\ 1\ 0\ 0\ 0\ 1\ 1\ 0\ 0\ 1\ 0\ 0\ 0 \\ 1\ 0\ 1\ 0\ 0\ 0\ 1\ 0\ 0\ 0\ 1\ 0\ 1\ 0\ 1\ 0\ 1\ 0\ 0\ 1\ 1\ 1\ 0\ 1\ 1\ 0\ 1\ 0\ 1\ 0 \\ 1\ 0\ 1\ 0\ 0\ 0\ 0\ 0\ 1\ 0\ 0\ 0\ 0\ 0\ 1\ 0\ 1\ 0\ 0\ 1\ 0\ 0\ 1\ 1\ 0\ 0\ 0\ 1\ 1\ 0 \\ 0\ 1\ 0\ 1\ 0\ 1\ 0\ 1\ 1\ 1\ 0\ 1\ 1\ 1\ 0\ 1\ 0\ 1\ 0\ 1\ 0\ 1\ 1\ 0\ 0\ 0\ 1\ 0\ 0\ 1 \\ 0\ 1\ 0\ 1\ 1\ 1\ 0\ 1\ 0\ 1\ 1\ 1\ 1\ 1\ 1\ 0\ 1\ 1\ 0\ 0\ 0\ 0\ 1\ 1\ 1\ 1\ 1\ 1\ 0 \end{bmatrix}.$$

Define a permutation π by the bijection function

π: (14 25 9 18 30 8 21 1 10 29 5 26 3 11 23 28 15 2 7 12 20 6 17 4 27 16 24 13 22 19)
→ (1 2 3 4 5 6 7 8 9 10 11 12 13 14 15 16 17 18 19 20 21 22 23 24 25 26 27 28 29 30)

Equivalently, let $R$ be the 30×30 permutation matrix defined by π. The public key is then

$$G = S(G_{PQ} + \tilde{G})R = \begin{bmatrix} 1\ 0\ 1\ 0\ 1\ 1\ 0\ 0\ 1\ 1\ 0\ 0\ 1\ 0\ 1\ 1\ 0\ 1\ 0\ 1\ 0\ 1\ 1\ 1\ 1\ 0\ 0\ 1\ 0\ 0 \\ 1\ 0\ 1\ 0\ 0\ 1\ 0\ 1\ 1\ 0\ 1\ 0\ 0\ 0\ 1\ 0\ 0\ 1\ 0\ 1\ 0\ 1\ 0\ 1\ 1\ 1\ 1\ 1\ 0\ 1 \\ 0\ 1\ 0\ 0\ 0\ 0\ 1\ 1\ 0\ 1\ 0\ 0\ 1\ 1\ 0\ 0\ 1\ 0\ 1\ 0\ 1\ 0\ 1\ 0\ 1\ 0\ 1\ 1\ 1\ 0 \\ 0\ 0\ 1\ 0\ 0\ 0\ 0\ 1\ 0\ 1\ 0\ 0\ 1\ 0\ 1\ 1\ 1\ 0\ 0\ 0\ 1\ 0\ 1\ 0\ 0\ 0\ 1\ 0\ 0\ 0 \\ 1\ 0\ 1\ 1\ 1\ 1\ 0\ 0\ 1\ 0\ 0\ 0\ 0\ 0\ 1\ 0\ 0\ 1\ 0\ 1\ 1\ 1\ 0\ 1\ 1\ 1\ 0\ 1\ 1\ 0 \\ 1\ 1\ 0\ 1\ 0\ 1\ 0\ 0\ 1\ 1\ 1\ 1\ 0\ 1\ 0\ 1\ 1\ 1\ 0\ 1\ 0\ 1\ 0\ 1\ 1\ 1\ 1\ 1\ 0\ 1 \end{bmatrix}.$$

Since $S$ is nonsingular and $R$ is a permutation, rank is preserved, so $G$ is full-rank.

For the message vector $\boldsymbol{m} = [1\ 1\ 1\ 0\ 0\ 1]$, the codeword $\boldsymbol{c}$ of the MCC corresponding to $G$ is given by

$$\boldsymbol{c} = \boldsymbol{m}G = [1\ 0\ 0\ 1\ 1\ 1\ 1\ 0\ 1\ 1\ 0\ 1\ 0\ 0\ 0\ 0\ 0\ 1\ 1\ 1\ 1\ 1\ 0\ 1\ 0\ 0\ 1\ 0\ 1\ 0].$$

Inject a random weight-3 error vector

$$\boldsymbol{e} = [0\ 0\ 0\ 1\ 0\ 0\ 0\ 0\ 0\ 0\ 0\ 0\ 0\ 0\ 0\ 0\ 1\ 0\ 1\ 0\ 0\ 0\ 0\ 0\ 0\ 0\ 0\ 0\ 0\ 0],$$

(i.e., errors at positions 4, 17,19). The received vector is given by

$$\boldsymbol{c_e} = \boldsymbol{c} + \boldsymbol{e} = [1\ 0\ 0\ \underline{0}\ 1\ 1\ 1\ 0\ 1\ 1\ 0\ 1\ 0\ 0\ 0\ 0\ \underline{1}\ 1\ \underline{0}\ 1\ 1\ 1\ 0\ 1\ 0\ 0\ 1\ 0\ 1\ 0],$$

where the erroneous bits are underlined.

### Decryption

**Step1** - **Inverse Permutation:**

Apply the inverse permutation π⁻¹ on the bits of $\boldsymbol{c_e}$:

π⁻¹: (8 18 13 24 11 22 19 6 3 9 14 20 28 1 17 26 23 4 30 21 7 29 15 27 2 12 25 16 10 5)
→ (1 2 3 4 5 6 7 8 9 10 11 12 13 14 15 16 17 18 19 20 21 22 23 24 25 26 27 28 29 30)

This operation can also be described by multiplying $\boldsymbol{c_e}$ on the right by an equivalent 30×30 permutation matrix $R^{-1}$:

$$\tilde{\boldsymbol{c}} = \boldsymbol{c_e} R^{-1} = [0\ 1\ 0\ 1\ 0\ 1\ \underline{0}\ 1\ 0\ 1\ 0\ 1\ 0\ 1\ \underline{1}\ 0\ 0\ \underline{0}\ 0\ 1\ 1\ 1\ 0\ 1\ 0\ 1\ 0\ 0\ 1\ 1].$$

**Step 2 - Unmasking:**

The set $\mathcal{M}$ of unmasked vectors is given by Equation (21) where

$$\mathrm{LS}(\mathcal{L}) = \{(1010\ldots10), (0101\ldots01), \mathbf{0}, \mathbf{1}\}$$

We now deinterleave the four vectors of $\mathcal{M}$ to their unmasked polynomial constitutes. There are only four possible unmasked variants:

$$(\tilde{\mathbf{c}} - \mathbf{0})_0 = x^7+x^{10}+x^{14}$$
$$(\tilde{\mathbf{c}} - \mathbf{1})_0 = 1+x+x^2+x^3+x^4+x^5+x^6+x^8+x^9+x^{11}+x^{12}+x^{13}$$
$$(\tilde{\mathbf{c}} - \mathbf{0})_1 = 1+x+x^2+x^3+x^4+x^5+x^6+x^9+x^{10}+x^{11}+x^{12}+x^{14}$$
$$(\tilde{\mathbf{c}} - \mathbf{1})_1 = x^7+x^8+x^{13}$$

**Step 3 - Inverting the High-Memory Polynomial Multiplication:**

Using Equation (22) and (23) we divide the first two outcomes of Step 2 by $\mathbf{q}_0$ and the last two by $\mathbf{q}_1$:

$(\mathbf{d}_0)_0 = (x^7+x^{10}+x^{14})/(1+x^7) = x^3+x^7$; remainder $= x^3$

$(\mathbf{d}_1)_0 = (1+x+x^2+x^3+x^4+x^5+x^6+x^8+x^9+x^{11}+x^{12}+x^{13})/(1+x^7) = x+x^2+x^4+x^5+x^6$; remainder $= 1+x^3$

$(\mathbf{d}_0)_1 = (1+x+x^2+x^3+x^4+x^5+x^6+x^9+x^{10}+x^{11}+x^{12}+x^{14})/x^7 = x^2+x^3+x^4+x^5+x^7$; remainder $= 1+x+x^2+x^3+x^4+x^5+x^6$

$(\mathbf{d}_1)_1 = (x^7+x^8+x^{13})/x^7 = 1+x+x^6$; remainder $= 0$

In vector form:

$$(\mathbf{d}_0)_0 = [0\ 0\ 0\ 1\ 0\ 0\ 0\ 1]$$
$$(\mathbf{d}_1)_0 = [0\ 1\ 1\ 0\ 1\ 1\ 1\ 0]$$
$$(\mathbf{d}_0)_1 = [0\ 0\ 1\ 1\ 1\ 1\ 0\ 1]$$
$$(\mathbf{d}_1)_1 = [1\ 1\ 0\ 0\ 0\ 0\ 1\ 0].$$

**Step 4 - Quotient Interleaving:**

The set $\mathfrak{D}$ four interleaved quotients is computed according to Equations (24) and (25):

$$\mathfrak{D} = \{\mathbf{d}_0, \mathbf{d}_1, \mathbf{d}_2, \mathbf{d}_3\}$$

where

$$\mathbf{d}_0 = (\mathbf{d}_0)_0 \curlywedge (\mathbf{d}_0)_1 = [0\ 0\ 0\ 0\ 0\ 1\ 1\ 1\ 0\ 1\ 0\ 1\ 0\ 0\ 1\ 1]$$
$$\mathbf{d}_1 = (\mathbf{d}_0)_0 \curlywedge (\mathbf{d}_1)_1 = [0\ 1\ 0\ 1\ 0\ 0\ 1\ 0\ 0\ 0\ 0\ 0\ 0\ 1\ 1\ 0]$$
$$\mathbf{d}_2 = (\mathbf{d}_1)_0 \curlywedge (\mathbf{d}_0)_1 = [0\ 0\ 1\ 0\ 1\ 1\ 0\ 1\ 1\ 1\ 1\ 1\ 1\ 0\ 0\ 1]$$
$$\mathbf{d}_3 = (\mathbf{d}_1)_0 \curlywedge (\mathbf{d}_1)_1 = [0\ 1\ 1\ 1\ 1\ 0\ 0\ 0\ 1\ 0\ 1\ 0\ 1\ 1\ 0\ 0]$$

**Step 5 - Parallel Viterbi Decoding:**

Decoding the four interleaved quotients in parallel yields the most likely codeword $\hat{\mathbf{d}}$ according to Equation (26), with the corresponding transformed plaintext $\hat{\mathbf{m}}_rS$. In this example, the CC has a free distance of $d_{free} = 5$, which implies that the code can reliably correct up to two errors within a sliding window of six bits. If three or more errors occur within such window, the parallel Viterbi decoding step is likely to fail in recovering the correct plaintext. Nevertheless, such decoding failure, can be detected by the CRC.

The trellis for the case $d_3 = (d_1)_0 \wedge (d_1)_1$ is depicted in Fig. 3. The highlighted path in the trellis—corresponding to the least-weight path—has an accumulated distance of 2 from $d_3$, which corresponds to the following information vector: (11011000). For all other Viterbi decoders, the most likely path through their respective trellises maintain a minimum distance greater than two from the corresponding $d_i$.

Note that for a CC with memory length 2, the final two zeroes result from forcing the trellis to converge to a single termination state. By discarding these appended zeroes, the most likely plaintext is obtained as:

$$\hat{m}S = [110110].$$

**Step 6** - **Plaintext Recovery:**

Using the inverse matrix

$$S^{-1} = \begin{bmatrix} 1 & 0 & 1 & 1 & 1 & 0 \\ 0 & 1 & 1 & 0 & 1 & 1 \\ 0 & 0 & 1 & 0 & 0 & 0 \\ 0 & 0 & 1 & 1 & 1 & 0 \\ 0 & 0 & 0 & 0 & 1 & 0 \\ 0 & 0 & 1 & 0 & 1 & 1 \end{bmatrix},$$

we can recover the original plaintext $m$ as follows:

$$m = \hat{m}SS^{-1} = [110110]S^{-1} = [111001]$$

Thus, the recovered plaintext is 111001, confirming correctness. In this example, the recovery of $m$ required only a single iteration. In the general case, the remainder resulting from the division of $\hat{m}_r(x)$ by $r(x)$ must be computed. If this remainder in non-zero, the selected codeword $\hat{d}$ should be discarded. The process must then be repeated iteratively with the next most likely candidate until a valid $d_i$ is identified or all possible candidates have been exhausted.

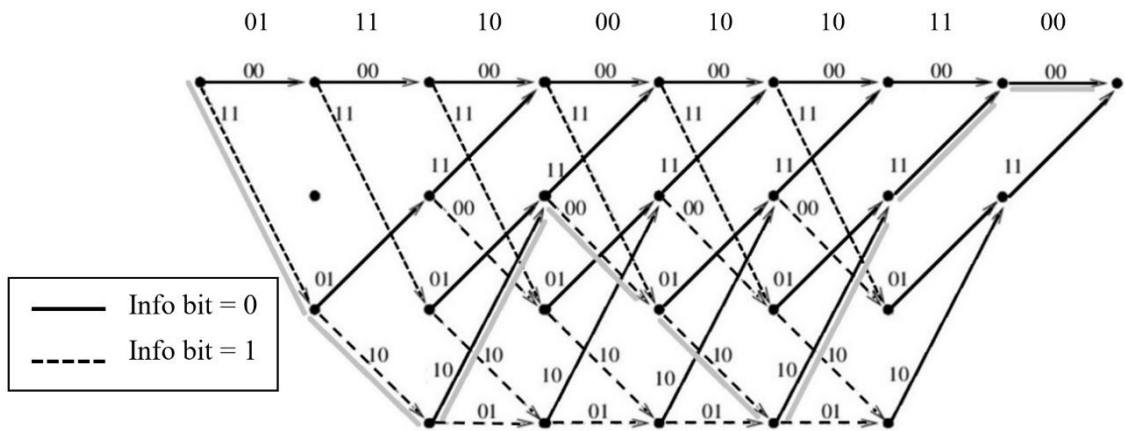

Figure 3. Trellis and least-weight path corresponding to $d_3 = [0\ 1\ 1\ 1\ 1\ 0\ 0\ 0\ 1\ 0\ 1\ 0\ 1\ 1\ 0\ 0]$

## 9. CRYPTANALYSIS RESISTANCE AND PERFORMANCE

In code-based cryptography, the complexity of cryptanalysis is primarily determined by the ISD algorithm, which targets the problem of decoding random linear block codes in the presence of errors. This task is considered computationally hard for large code dimensions and high error weights. However, if the public code deviates from a random code due to hidden structures, ISD may not reflect the true attack complexity. The underlying mathematical problem is the syndrome decoding problem. In our context, the finite-dimensional matrix $G$, which serves as the generator matrix for the MCC (allowing it to be treated as a linear block code), has an associated parity-check matrix $H$ satisfying:

$$GH^T = \mathbf{0} \tag{36}$$

Given a syndrome $s$, and assuming that exactly $t$ errors have occurred, the challenge is to find an error vector $e$ such that

$$He^T = s \quad \text{and} \quad \text{wt}(e) = t \tag{37}$$

The cryptanalysis complexity is estimated using the best-known ISD algorithms (e.g., Prange, Lee-Brickell, BJMM) [8-10], which assess security as bit operations, often ignoring memory usage or parallelization. These estimates assume a fixed weight wt($e$). However, in our method, the exact error count is unknown—only the error probability $e$ is known—introducing additional complexity compared to other code-based schemes. Furthermore, while structured error patterns may affect complexity, the MCC's error vector $e$ is truly random and, by definition, unstructured.

In the context of quantum computing, ISD complexity is typically estimated based on classical assumptions. Although quantum algorithms like Grover's [11] can offer quadratic speedup, no exponential improvement in ISD is known. Therefore, ISD-based complexity assessments in code-based cryptography assume ISD remains the most efficient attack.

To compare the complexity of MCC with other code-based methods, we adopt the complexity measure described in [11]. Specifically, for a classical computer attacking of an ($N$, $K$, $t$) code, the ISD complexity—denoted $C_{\text{ISD}}$—is proportional to the number of iterations performed by the ISD algorithm. In our notation, we have

$$C_{\text{ISD}} \sim \frac{\binom{N}{K}}{0.29\binom{N-t}{K}} \tag{38}$$

and for quantum computing, the complexity, denoted $C_{\text{QISD}}$, is given by

$$C_{\text{QISD}} \sim \sqrt{\frac{\binom{N}{K}}{0.29\binom{N-t}{K}}} \tag{39}$$

where each quantum iteration involves a function evaluation requiring $O(N^3)$ qubit operations.

We now apply Equation (38) to evaluate the complexity of a Goppa-code variant, specifically the (4096, 3556, 45) code, and compare it with that of an MCC offering similar key lengths. Notably, MCC provides an additional layer of security: while the parameter $e$ is publicly known, the actual number of errors $t$ remains unknown. Furthermore, the decoder's use of polynomial division introduces additional, unpredictable errors, further obscuring the effective error count. Consequently, whereas ISD algorithms assume a fixed value of $t$, decoding in MCC entails iterating over a range of possible error weights. For a meaningful comparison, however, we assume that in the MCC scheme, the effective number of errors is fixed at $eN + \alpha$, where $\alpha$ represents the additional errors introduced during polynomial division, as defined in Equation (32). Substituting the parameters of the Goppa (4096, 3556, 45) code we obtain:

$$C_{\text{ISD}}(\text{Goppa}) \approx \frac{\binom{4096}{3556}}{0.29\binom{4096-45}{3556}} \approx 7.06 \times 10^{40} \qquad (40)$$

For an MCC with a comparable key length, we can choose parameters $N = 5600$ and $K = 2600$. For instance, by selecting polynomials $\boldsymbol{p}_0(x)$ and $\boldsymbol{p}_1(x)$ of degree 14 and, $\boldsymbol{q}_0(x)$ and $\boldsymbol{q}_1(x)$ of degree 186 (resulting in a rate ½ MCC with memory length 200 and 2600 input bits) the corresponding 5600-bit codeword achieves a public key length comparable to Goppa code, approximately 2 megabytes in both cases. For these parameters, selecting $\boldsymbol{q}_0(x) = x^{93}$ and $\boldsymbol{q}_1(x) = 1+x^{186}$ with $e = 0.02$ yield an error probability at the Viterbi decoder input of:

$$\frac{eN + \alpha}{N} \approx 0.07$$

This corresponds to an average of approximately 392 errors in a 5600-bit ciphertext. Such an error rate lies well within the correction capability of a rate ½ CC with memory length 14, making decoding failures highly improbable. By contrast, incorrectly unmasked candidates are decoded (as analysed in Section 7) to codewords lying at a Hamming distance of approximately

$$0.11 \times 5600 \approx 616$$

from the received vector. This distance exceeds the expected correction radius of 392 errors by a wide margin, placing such candidates outside the decoder's error sphere and thereby making them easily distinguishable from the correct solution. The separation implied by the code's distance spectrum ensures that the valid plaintext can be identified with overwhelming probability, while incorrect candidates are systematically rejected. Under these assumptions, the decryption complexity of the MCC is:

$$C_{\text{ISD}}(\text{MCC}) \approx \frac{\binom{5600}{2600}}{0.29\binom{5600-392}{2600}} \approx 2.088 \times 10^{112} \qquad (41)$$

This result reflects a security improvement by a factor of approximately $0.295 \times 10^{62} \approx 2^{204}$ compared to the Classic McEliece scheme in Equation (40). For quantum ISD (QISD), the improvement is roughly $2^{102}$, consistent with the square-root speedup provided by Grover's algorithm.

Increasing the public key length further enhances resistance to cryptanalysis, as indicated by Equation (38). For example, compare the most practical (8192, 6528, 128) binary Goppa code used in Classic McEliece with the CC parameters of the previous example, where

$$\boldsymbol{q}_0(x) = x^{193},\ \boldsymbol{q}_1(x) = 1+x^{386},\ K = 5000,\ N = 10800.$$

Selecting an effective error rate of $e = 0.017$ yields an error probability at the Viterbi decoder of approximately 0.06, corresponding to an average of about 648 errors in a 10800-bit ciphertext.

Substituting the respective parameters for the two codes into Equation (38) gives a security improvement of the MCC over Classic McEliece by a factor of approximately $1.6 \times 2^{306}$.

Since the primary purpose of the MCC cryptosystem is to securely distribute a session key for subsequent reuse, a public-key size of several megabytes is not considered a practical limitation—except in cases where the transmitted data message itself is extremely short.

To estimate the probability of a decoding failure (and thus the need for retransmission), consider the following example. Suppose we employ a rate ¼ CC with memory length 10, defined by:

$$G_P(x) = [2327, 2313, 2671, 3175] \qquad (42)$$

where the polynomials are expressed in octal form. This code provides robust error correction capabilities, characterised by a free distance of $d_{free} = 29$. In practical terms, this means that the Hamming distance between any two paths through the 1024-state trellis—which diverge from a common state and remerge after 11 segments (corresponding to 44 bits)—is at least 29. Consequently, the CC can reliably correct up to 14 errors within each 44-bit window.

Next consider the following high-memory polynomials:

$$G_Q(x) = [1+x^{495}+x^{990}, x^{247}, x^{743}, 1+x^{990}].$$

For an error rate $e = 0.04$, simulations yield

$$\frac{\alpha}{N} \approx \frac{0.18 + 0 + 0 + 0.13}{4} \approx 0.0775$$

Thus

$$e + \frac{\alpha}{N} = 0.04 + 0.0775 = 0.1175$$

The Viterbi decoder fails if more than 14 errors occur within a 44-bit window. At an effective error rate of 0.1175, the probability of such an event is approximately $8.998 \times 10^{-5}$. Over 228 windows (a ciphertext length of approximately 10,000 bits) the probability of successful decoding is therefore

$$(1 - 8.998 \times 10^{-5})^{228} \approx 0.98$$

This result demonstrates that the MCC cryptosystem maintains a high decryption success rate even for ciphertexts spanning tens of thousands of bits.

We note that low-rate public codes can correct a larger fraction of errors than high-rate codes, but their higher redundancy can impact both the efficiency and security of code-based cryptosystems, as indicated by Equation (38). When $K \ll N$, the large number of parity constraints $N - K$ relative to the small number $K$ of message bits may be leveraged by an attacker to locate low-weight codewords or speed up decoding, as shown in [12]. Consequently, the code rate should be a key factor when choosing the underlying CC construction.

Finally, we examine the computational complexity of decryption. Schöffel *et al.* [13] present a hardware/software co-design implementation of a Hamming Quasi-Cyclic (HQC) cryptosystem tailored for IoT edge devices, providing a detailed evaluation of energy consumption, performance, and deployment trade-offs. The study demonstrates that code-based schemes can constitute practical alternatives to lattice-based approaches in resource-constrained environments. In [14], another hardware accelerator for HQC-based post-quantum cryptography is introduced, implementing key generation, encapsulation, and decapsulation on an FPGA with substantial performance gains achieved through tight integration with a RISC-V core.

In the proposed MCC cryptosystem, the primary contributor to decryption complexity is the Viterbi decoding stage. Other operations—such as polynomial division, unmasking, interleaving, and deinterleaving—are applied to the entire received ciphertext but incur a negligible computational cost compared to the per-bit processing requirements of the Viterbi algorithm. To quantify this cost, we evaluate the number of Add-Compare-Select (ACS) modules utilized per decrypted plaintext bit. These modules can be efficiently realized in both software and hardware, enabling practical and scalable deployment.

Suppose the MCC employs a CC with memory $p$ and rank-$l$ masking matrix. In this case the decryption process requires $2^l$ parallel Viterbi decoders, each utilizing $2^p$ ACS modules per bit. Consequently, the total number of ACS modules per plaintext bit is $2^{l+p}$. This complexity measure scales linearly with the plaintext length $K$, ensuring predictable and manageable performance as the data size grows. For example, using the CC defined in Equation (42) with $l = 5$ and memory $p = 10$, the total ACS operations per bit become $2^{l+p} = 2^{15}$, a computational load well within the capabilities of commercial processors, including those found in modern mobile devices.

## 10. CONCLUSION

This work presents a novel post-quantum encryption scheme based on noise-enhanced high-memory convolutional codes, addressing fundamental limitations of traditional block-code-based cryptosystems. Security is reinforced through semi-invertible transformations that generate fully dense, random-like generator matrices, thereby eliminating vulnerabilities associated with low-density or structured codes.

The proposed design combines high-rate intentional noise injection with additional stochastic noise inherently produced by polynomial division during decryption. This layered mechanism substantially strengthens resistance to cryptanalysis, even under the conservative assumption that an adversary has complete knowledge of the underlying convolutional code. While legitimate recipients resolve decryption ambiguity in polynomial time, adversaries confront exponential-time complexity. The interplay among semi-invertible masking, polynomial division, and invertible scrambling ensures that decryption without the private key remains computationally infeasible.

Empirical analysis indicates that the proposed scheme achieves security margins exceeding those of Classic McEliece—by factors greater than $2^{100}$ against quantum adversaries and $2^{200}$ against classical adversaries—while maintaining efficient decoding performance.

Beyond its strong security guarantees, the scheme supports arbitrary plaintext lengths, provides linear scalability, and enables high-throughput implementation through parallel directed-graph decoders. Collectively, these properties make the proposed construction a robust, scalable, and practically deployable candidate for next-generation post-quantum public-key cryptography.

## REFERENCES

[1]   McEliece, R.J., (1978) "A Public-Key Cryptosystem Based on Algebraic Coding Theory", DSN Progress Report.
[2]   Bernstein, D.J., Lange, T., Peters, C., (2008) "Attacking and defending the McEliece cryptosystem", Post-Quantum Cryptography, pp. 31–46.

[3]     Misoczki, R.; Tillich, J.-P.; Sendrier, N.; Barreto, P.S.L.M., (2013) New McEliece Variants from Moderate Density Parity-Check Codes", Proceedings of the IEEE International Symposium on Information Theory (ISIT), pp. 2069–2073.

[4]     NIST Post-Quantum Cryptography Standardization, (2023) "NIST Post-Quantum Cryptography Standardization", https://csrc.nist.gov

[5]     Weger, V.; Barelli, E.; Beullens, W.; Couvreur, A.; Debris-Alazard, T.; Galvez, L.; Gauthier-Umaña, V.; Misoczki, R.; Puchinger, S.; Sárközy, G.; Sendrier, N., (2022) "A Survey on Code-Based cryptography", IACR Cryptology ePrint Archive, Report 2022/005. Updated July 2024, https://eprint.iacr.org/2022/005

[6]     Cayrel, P.L.; Gaborit, P.; Misoczki, R.; Puchinger, S.; Weger, V. (eds.), (2024) Code-Based Cryptography, Proceedings of CBCrypto 2024, Vol. 14699, Springer, Cham.

[7]     Baldi, M.; Chiaraluce, F.; Garello, R.; Mininni, F., (2007) "Quasi-cyclic Low-Density Parity-Check Codes in the McEliece Cryptosystem", Proceedings of the IEEE International Conference on Communications (ICC), pp. 951–956.

[8]     Becker, A.; Joux, A.; May, A.; Meurer, A., (2012) "Decoding Random Binary Linear Codes in $2^{n/20}$", Advances in Cryptology – EUROCRYPT 2012, Lecture Notes in Computer Science Vol. 7237, pages 520–536.

[9]     Prange, E., (1962) "The Use of Information Sets in Decoding Cyclic Codes", IRE Transactions on Information Theory, Vol. 8, No. 5, pp. 5–9.

[10]    Lee, P.J.; Brickell, E.F., (1988) "An Observation on the Security of McEliece's Public-Key Cryptosystem", Advances in Cryptology – EUROCRYPT 88, LNCS Vol. 330, pp. 275–280.

[11]    Bernstein, D.J., (2010) "Grover vs. McEliece", in Sendrier, N. (ed.), Post-Quantum Cryptography, PQCrypto 2010, Vol. 6061, pp. 73–80. Springer, Heidelberg.

[12]    Carrier, N.; Espitau, T.; Gaborit, P.; Géraud, A.; Kachigar, G., (2024) "Reduction from Sparse LPN to LPN, Dual Attack 3.0", Advances in Cryptology – EUROCRYPT 2024, Vol. 14655, pp. 735–765. Springer, Cham.

[13]    Schöffel, P.; Feldmann, A.; Wehn, N., (2023) "Code-Based Cryptography in IoT: A HW/SW Co-Design of HQC", Proceedings of the Design, Automation & Test in Europe Conference, pp. 368–373.

[14]    Di Matteo, A.; Marrone, M.; Valenzano, A., (2025) "TYRCA: A RISC-V Tightly Coupled Accelerator for Code-Based Cryptography", Proceedings of the 62nd Design Automation Conference (DAC 2025), pp. 1–6.

[15]    M. Ariel, (2025) "High-Memory Masked Convolutional Codes for Post-Quantum Cryptography", CS & IT Conference Proceedings (2025) 15 (17), pp. 1–22.

**AUTHOR**

**Meir Ariel** received his B.Sc. and M.Sc. degrees (with honours) in Electrical Engineering and a Ph.D. in Algebraic Group Theory, all from Tel Aviv University. He has over 30 years of R&D experience in signal processing, wireless communications, and space technologies, with leadership roles across industry and the public sector. From 1999 to 2013, he was CEO of technology ventures in vision systems, VoIP, and fintech. Since 2013, he has served as Director General of the Herzliya Science Center, and in 2018, he founded Tel Aviv University's Space Engineering Center, leading 23 nanosatellite developments for space research and experimental communications. He holds 15 patents in information theory and signal processing and has served on national advisory boards in science and education. 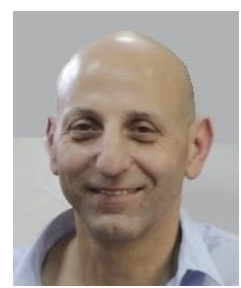
His honours include the Marco Polo Society Award (2018) and recognition by Israel's Ministry of Science and Technology as one of the country's 60 pioneering inventors (2016).